\documentclass{Odyssey2026}

\interspeechcameraready 

\title{I Hear, Therefore I Trust: A Socio-Technical Investigation of Humans as Synthetic Speech Detectors}
\author[affiliation={1}]{Lelia}{Erscoi}
\author[affiliation={1}]{Tomi}{Kinnunen}

\affiliation{Computational Speech Group}{University of Eastern Finland}{Joensuu, Finland}
\email{lerscoi@uef.fi}

\usepackage{comment}
\usepackage{soul}
\usepackage{hyperref}
\usepackage{acro}
\usepackage{svg}
\usepackage{multirow}
\usepackage{appendix}
\usepackage{float}

\begin{document}

\maketitle
\begin{abstract}
    
   Automatic deepfake detection has received considerable research attention, yet the socio-technical environment in which humans actually encounter synthetic speech remains poorly understood. We investigate voice deepfake detection as a perceptual and contextual process, presenting a localization task in which 47 participants marked suspected synthetic segments across authentic, fully synthetic, and partially synthetic utterances under three manipulated trust cues: instructional framing, affective priming, and provenance labeling. Participants provided quality ratings on mechanicalness, expressiveness, intelligibility, clarity, calmness, and confidence of evaluation. Utterance class was the primary determinant of detection accuracy and perceptual quality; trust cues produced no main effects but motivated detection behavior. Fully synthetic speech was detected at below-chance levels. Quality ratings tracked utterance type, indicating implicit discrimination where overt detection failed. 
    
\end{abstract}

\section{Introduction}
\label{sec:intro}
Generative artificial intelligence (genAI), particularly conversational agents, extends a socio-cultural transition of epistemic authority and information sourcing, from traditional news outlets, to social media platforms and now to personalized chatbots \cite{johansen2020public, nehring2024large}. Within this shifting landscape, trust becomes a currency of the attention economy \cite{myers2025attention}. Communities such as Reddit's r/isthisAI\footnote{https://www.reddit.com/r/isthisAI/} have millions of weekly visitors who propose content for open debate on authenticity. The ease of creation and distribution allows for an overflow of synthetic media content that dampens the user detection capabilities and creates a liar's dividend, where anything encountered online might plausibly be synthetic \cite{harris2024synthetic}. In this study, we aim to draw attention to speech deepfakes as a socio-technical trust issue \cite{laas2023deepfakes} that remains under-investigated, particularly given that human detection performance is typically studied in isolation from the broader stream of cues listeners encounter in everyday communicative settings. Even so, detection accuracy tends to be limited \cite{warren2024better, alali2025partial, bhatti2026can}, raising alarms about user vulnerabilities in the face of increasingly believable generative models.

Speech is especially susceptible to misinterpretation when presented in isolation from other cues that could support holistic authenticity assessment. Beyond serving as a carrier for linguistic content, the human voice occupies a unique position in trust judgments, with paralinguistic dimensions (e.g., tone, pitch, intonation) additionally carrying associations with identity, intent, familiarity, or authenticity \cite{fan2025machines}. Going beyond traditional voice spoofing attacks (or presentation attacks) targeted against automatic speaker verification systems \cite{wu2015spoofing}, voice deepfakes blend into the broader content stream, where credibility is inferred from arbitrary metrics such as number of likes or even personal attitudes towards the content \cite{buchanan2020people}. In such contexts, automatic detection systems or simple disclosure labels may be insufficient, since warnings compete within the same attention economy \cite{harris2024ai}. When reproducibility and ease of dissemination are no longer technical constraints, the risk of voice deepfakes lies in how they are received, as well as in how trust is manufactured in everyday communicative settings \cite{chiou2020we,scott2020human}.

The rapid maturation of speech synthesis technology \cite{casanova2022yourtts} tends to outpace countermeasure development in an arms race fueled by ever-growing data availability. On the automatic detection front, performance evaluation benchmarks such as the ASVspoof challenge series \cite{liu2023asvspoof, wang2024asvspoof} indicate strong synthetic speech detection accuracy, especially under controlled conditions. Contrastingly, synthetic speech detection accuracy in naive, unattentive listeners hovers near chance levels \cite{warren2024better,alali2025partial}. More alarmingly, listeners tend to rely on outdated acoustic cues such as robotic prosody or abnormal pausing and breathing, elements that modern synthesis systems no longer reliably produce \cite{warren2024better,amirkhani2025detecting}. This leaves judgments vulnerable to the very contextual and social factors that have received little empirical attention. When only segments of an utterance are manipulated, detection rates fall even among attentive listeners \cite{alali2025partial}.

The present study moves beyond purely computational concerns to examine the broader socio-technical phenomena of voice deepfake detection. Building on previous frameworks describing trust and artificial intelligence \cite{glikson2020human, lockey2021review} or deepfakes specifically \cite{laas2023deepfakes}, we focus on the environmental context as the critical variable often overlooked in controlled lab settings. Crucially, no prior study in human deepfake detection has jointly manipulated the acoustic manipulation type, the contextual trust environment, and the listener response modality within a single ecologically valid protocol. This fragmentation means that existing benchmarks are incomplete by design, that they measure detection under conditions that do not reflect how people actually encounter synthetic speech. The result is a systematic overestimate of human detection capability.  We argue that neither automatic detection systems nor human listeners alone offer a silver bullet, as both are subject to distinct biases \cite{harris2024ai, amirkhani2025detecting, mustak2023deepfakes,  li2025we, somoray2025human}. Consequently, effective mitigation requires understanding the interaction between these agents and the contextual factors that shape their performance. 

This paper offers three primary contributions. First, we delineate the environmental context as a distinct dimension of the socio-technical trust framework, identifying specific sub-sources of trust (such as instructional framing, affective priming, and provenance labels) that mediate human susceptibility to voice deepfakes. Second, we systematically investigate how these contextual factors influence detection accuracy and perceptual quality, testing four hypotheses: if detection accuracy varies by utterance authenticity (\textbf{H1}) and whether this relationship is moderated by trust orientation (\textbf{H1a}), and if subjective ratings vary by authenticity (\textbf{H2}) and trust orientation (\textbf{H2a}). Third, to advance ecological validity, we introduce a synthetic speech localization task. Unlike binary classification tasks, this allows participants to highlight specific segments they deem inauthentic, providing fine-grained insight into the decision-making process and revealing exactly where and why trust fractures in real-world listening scenarios.

\section{Socio-technical background}
The term "deepfake" was coined in the mid-2010s to describe non-consensual impersonating pornographic synthetic media produced with deep learning techniques \cite{bezio2018ctrl}. Over time, its definition has expanded to encompass generative techniques across audio, text, images, videos, and multimodal media. Beneficial applications such as language dubbing \cite{banos2023key} or personalized voice assistants \cite{pan2025grooming} have fueled public interest in generative AI, leading to a diminished perceived gravity and amplified potential for confusion and harm \cite{ cheng2025sycophantic}. When deepfakes disseminate disinformation or cause harm, culpability becomes increasingly diffused, with responsibility often shifted onto users who are framed as insufficiently skeptical \cite{harris2024synthetic}. Initiatives such as the European Union's Regulation 2024/1689 (Artificial Intelligence Act) \cite{eu2024aiact} and its corresponding AI Office\footnote{\href{European AI Office}{https://digital-strategy.ec.europa.eu/en/policies/ai-office}} aim to address AI technologies based on their assigned risk, with deepfakes categorized as a "transparency" risk.

\subsection{Trust and synthetic speech}
The issue of harm by (speech) deepfakes is not a matter of \emph{how}, but \emph{when} \cite{harris2024synthetic, li2025we,  khan2023battling}. With the normalization of synthetic media, increasingly realistic synthesis models create an expanding security gap at both societal and technical levels, where subtle alterations of reality threaten privacy, safety, and even the integrity of democratic processes \cite {harris2024synthetic}.

Mounting a spoofing attack has become increasingly straightforward, driven by open-access Text-to-Speech (TTS) models and the broad availability of voice recordings \cite{krikheli2026s}, including those that users willingly provide to build personalized voice-based agents \cite{krikheli2026s}. Automatic detection models often overlook cues that human listeners naturally pick up, such as emotional affect \cite{mahapatra2025can} or paralinguistic artifacts \cite{layton2025every}. For detection tools to function effectively as decision-support mechanisms, however, they must first establish cognitive trust with their users \cite{glikson2020human, chiou2020we}. In the absence of such trust, users are prone to disregard model outputs entirely. How to optimally combine human and machine capabilities, therefore, remains an open question, particularly given evidence that even minor inconsistencies in algorithmic outputs erode the very trust that makes human-AI collaboration effective \cite{glikson2020human}.

\subsection{Human perception and the limits of vigilance}
It remains an open question how realistic it is to expect individuals to be constantly vigilant. As with other technological security threats, such as malware, computational safeguards must be complemented by awareness and education, particularly since many people may still (mistakenly) believe that AI is "not quite there yet". However, as personal life increasingly merges with the digital domain (see, for example, digital personas \cite{de2013digital}), it becomes easier to tailor attacks that can deceive even trained professionals, especially when the impersonation relies solely on voice. In real-world conditions, spoofing attacks always come with intent and meaning. Attackers exploit trust by imitating a family member\footnote{https://people.com/woman-conned-out-of-usd15k-after-ai-cloned-daughters-voice-terrifying-scam-11775622} or by invoking authority to sway decisions or influence elections\footnote{https://f24.my/A6cV}. Moreover, synthetic voice generation need not arise from explicitly malicious motives to result in deception and harm, as illustrated by cases in which individuals develop romantic attachments to personalized chatbots, sometimes leaving their actual families \footnote{https://people.com/man-proposed-to-his-ai-chatbot-girlfriend-11757334} or even ending their own lives \footnote{https://edition.cnn.com/2025/11/06/us/openai-chatgpt-suicide-lawsuit-invs-vis}. 

Partial manipulation (interweaving synthetic bits within otherwise authentic content) requires far less technical overhead while achieving a significant impact through harvesting the quality and authority of the original signal. The challenge of discriminating authentic from synthetic speech may be contaminated by the perception of the whole, in which authentic regions anchor judgment and suppress suspicion even when manipulation is present, or vice versa, where the presence of manipulation discredits the overall credibility. Empirical evidence favors the latter as detection rates fall substantially when authentic and synthetic speech are intermixed \cite{alali2025partial}. This suggests that partial manipulation is not merely harder to detect but qualitatively different as a perceptual problem. 

\subsection{Trust cues as experimental manipulations}
Critically, trust in what one hears is not formed on acoustic evidence alone. The present study aims to investigate this multi-source account of trust through three manipulations, hereby referred to as \emph{trust cues}, working at the environmental level of the trust framework introduced in Section~\ref{sec:intro}. These cues each target a distinct channel through which contextual information reaches the listener, thereby supplementing their decision process on top of the acoustic properties of the utterance to be evaluated, whether synthetic, authentic, or a mix of the two.

\begin{itemize}
    \item \textbf{Instruction framing} (I$+$ / I$-$) manipulates the cognitive set with which participants approach the task. Participants are presented with one of two scenarios: one framing the audio as originating from a groundbreaking communication technology under positive evaluation (I$+$), the other framing it as material from a nefarious actor targeting platform users (I$-$). In both conditions, participants assumed the role of a social media platform moderator whose decisions carry real consequences for the platform's users, and received identical task instructions that remained visible throughout the experiment. This manipulation targets the top-down expectations that govern how attentively and skeptically a listener engages with incoming audio \cite{dall2014rating}.

    \item \textbf{Affective valence priming} (V$+$ / V$-$) manipulates the emotional context preceding each trial through images depicting positive- or negative-valence content. Drawing on evidence that affect modulates perceptual judgment \cite{loewenstein2003role, mittal2020emotions}, this manipulation tests whether emotional state at the moment of listening shifts detection sensitivity or evaluative bias.

    \item \textbf{Provenance labeling} (P$+$ / P$-$) manipulates credibility in source attribution by presenting a textual label affirming a trusted origin for the audio, with the negative trust option having no such label. This tests whether an explicit source claim overrides or interacts with acoustic evidence in forming authenticity judgments \cite{gamage2025labeling}. Provenance labels of this kind are increasingly being adopted in practice, and are set to become a regulatory requirement under the EU AI Act, with a dedicated Code of Practice on marking and labeling of AI-generated content currently being drafted ahead of the transparency obligations coming into force in 2026 \footnote{\href{Code of Practice on marking and labelling of AI-generated content}{https://digital-strategy.ec.europa.eu/en/policies/code-practice-ai-generated-content}}.
\end{itemize}

\section{Methods}
\subsection{Dataset}
\label{sec:dataset}
The LlamaPartialSpoof \cite{luong2025llamapartialspoof} is a recent dataset that contains both fully synthetic and partially spoofed speech. It contains English utterances from 40 LibriTTS \cite{zen2019libritts} speakers, with synthetic counterparts generated using five open-source models (LJ JETS\footnote{https://github.com/espnet/espnet/tree/master/egs2/ljspeech/tts1}, YourTTS\footnote{https://github.com/coqu$I-$ai/TTS/}, XTTS V2\footnote{https://huggingface.co/coqui/XTTS-v2}, GPT-SoVITS\footnote{https://huggingface.co/lj1995/GPT-SoVITS/tree/main}, CosyVoice\footnote{https://github.com/FunAudioLLM/CosyVoice}) plus one commercial service (ElevenLabs\footnote{https://elevenlabs.io}). Partially synthetic samples are created by splicing genuine and fake segments using cross-fading techniques, simulating realistic localized manipulation attacks. 

For the task at hand, 20 utterances were sampled in a ratio of 1:1:2 of authentic-synthetic-partially synthetic speech by balancing the generation model and sentence spoken. To maintain a manageable experiment duration and minimize participant listening fatigue \cite{zhang2018understanding}, utterances were limited to those over 10 seconds, and for partially synthetic speech, only utterances with a maximum of 3 synthetic sections were chosen. Files were then mixed with randomly selected environmental recordings from the International Soundscape Database \cite{mitchell2021international} at 25 dBSNR to introduce ecologically valid acoustic conditions reflecting real-world listening scenarios \cite{bockstael2018presenting}.

For the affective priming manipulation, images were drawn from the Open Affective Standardized Image Set (OASIS) \cite{kurdi2017introducing}, a standardized dataset containing user-validated arousal (calm-intense) and valence (negative-positive) ratings for each image. Images rated in the top 20\% for arousal and at either valence extreme were selected to maximize priming effects, with manual verification to exclude explicitly graphic content.

\subsection{Synthetic speech detection}
Participants were asked to evaluate the 20 utterances presented in a randomized order to prevent carryover effects \cite{wehrman2020decisional}. For each clip, participants are instructed to listen to the audio in its entirety and to identify any portions that sound artificial or fabricated. Participants could annotate the suspicious segments using two types of markers: point-based "flags" (timestamps) or interval-based "segments" (start-end ranges). Participants were provided unlimited time to attend to each utterance and instead were instructed to perform to the best of their ability.

\begin{figure}[!h]
    \centering
    \includegraphics[width=1\linewidth]{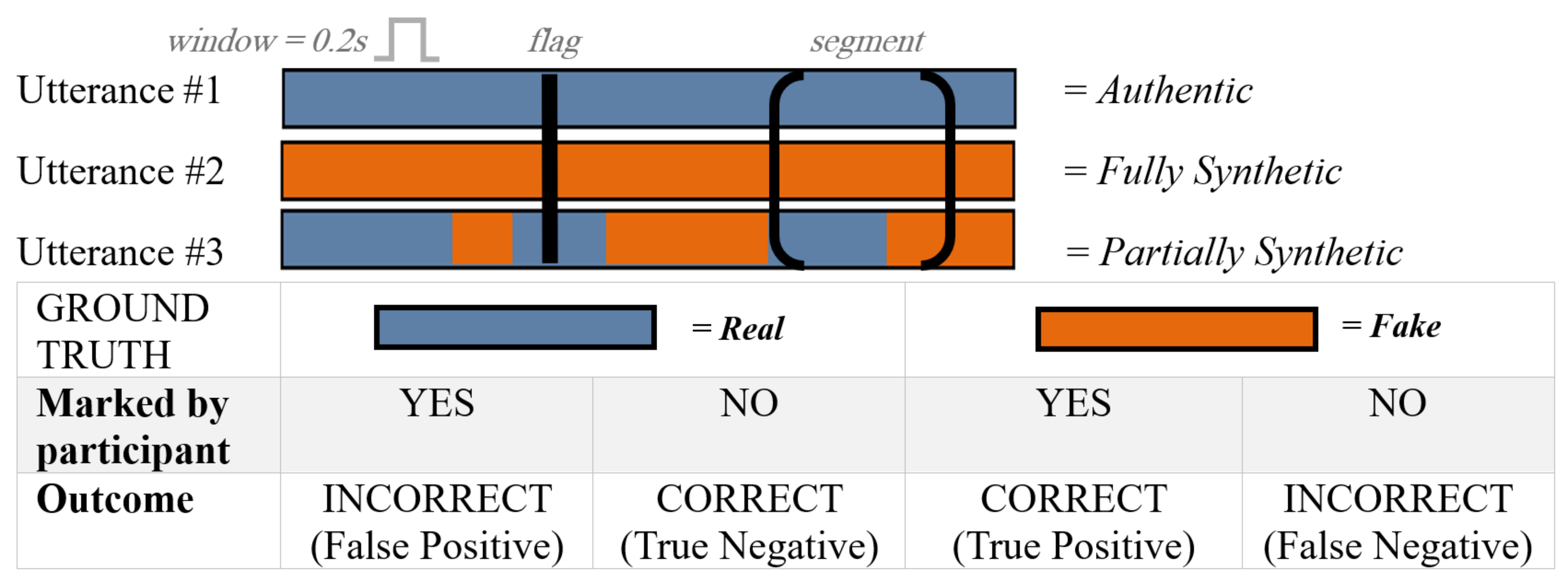}
    \caption{How user markers translate to evaluation metrics. Each speech excerpt is divided into windows (length $= 0.2s$) to extract the ground-truth label and to verify whether the user placed a marker (flag or segment) within that window. A $\pm200ms$ temporal margin is applied to each marker to account for reaction-time delays. }
    \label{fig:Markers}
    \vspace{-0.4cm}
\end{figure}

\subsection{Speech evaluation}
Participants also completed a set of subjective evaluation statements. For our study, each utterance was treated as a single perceptual entity by participants, regardless of any insertions or edits, ensuring that their judgments reflected the overall impression of the voice rather than isolated segments. The following dimensions were assessed on a five-point Likert scale (\textit{Strongly Disagree} $=-2$, \textit{Disagree} $= -1$, \textit{Unsure} $= 0$, \textit{Agree} $= +1$, \textit{Strongly Agree} $= +2$), with \textit{Unsure} as the default response. 
\begin{itemize}
    \item \textbf{Artificiality ("The voice sounds mechanical")}: Listeners are known to be sensitive to prosodic and timing cues that signal non-human origin regardless of prior task experience \cite{warren2024better}. This dimension deliberately targets markers of \emph{artificiality} rather than the more ambiguous concept of \emph{naturalness} \cite{cooper2025good}.

    \item \textbf{Expressiveness ("The voice sounds expressive")}: captures emotional variation and prosodic richness \cite{san2025human}. Synthetic voices may be perfectly intelligible yet lack the affective modulation characteristic of human communication \cite{fan2025machines}.

    \item \textbf{Intelligibility ("The voice is easy to understand")}: reflects the ease with which listeners comprehend the speech signal, affecting cognitive load and user acceptance \cite{ma2024intelligibility}.

    \item \textbf{Clarity ("The audio sounds clean")}: captures technical recording or synthesis quality, including compression artifacts or signal anomalies \cite{sokol2022automatic}. 

    \item \textbf{Calmness ("The voice sounds calm")}: assesses the perceived 
    arousal and emotional valence of the speaker. In deception contexts, voice-based credibility judgments are known to be influenced by perceived speaker affect \cite{inproceedingstrust, inproceedingsemotion}.

    \item \textbf{Confidence ("I am confident in my evaluation")}: captures subjective certainty about a judgment. It is well known that confidence does not reliably correlate with detection accuracy, and synthetic speech can remain highly deceptive even to confident listeners \cite{hashmi2024unmasking}. Confidence ratings map to detection behavior: marking no segment implies an authentic classification, meaning high confidence reflects certainty in non-detection rather than abstention.

\end{itemize}
The evaluation questions were presented in randomized order in each trial to avoid automaticity \cite{toner2015perils}.

\subsection{Task design}
\label{sec:taskdesign}

\begin{figure*}[h!]
    \centering
    \includegraphics[width=1\textwidth]{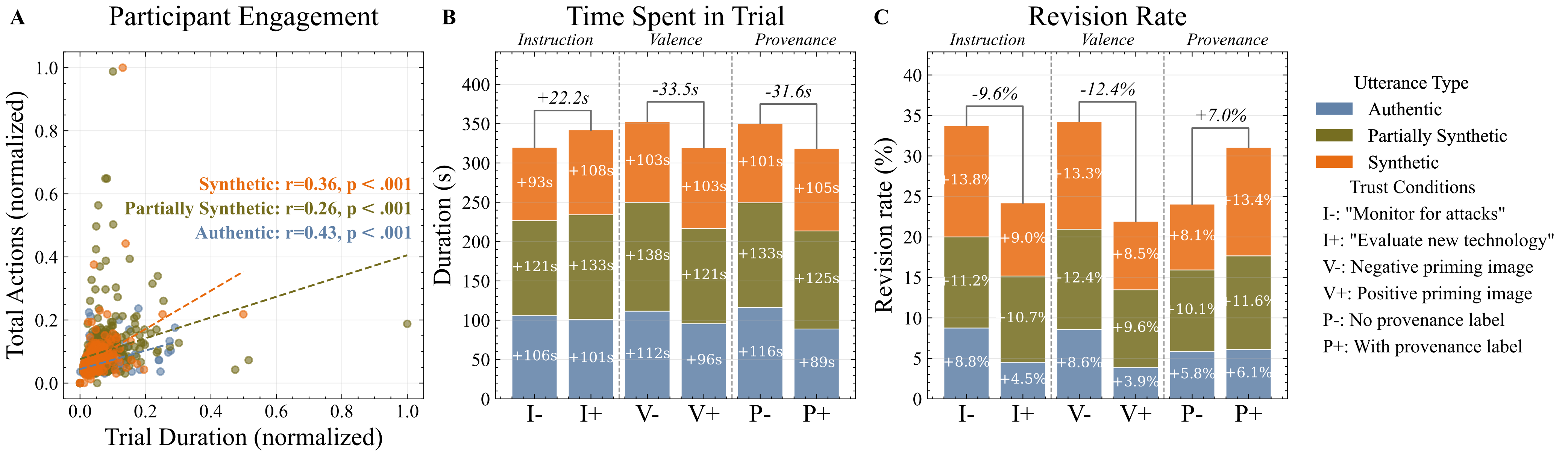}
    \caption{Decision-making patterns across conditions. (A) Positive correlation between action count and trial duration across all utterance types; synthetic speech trials required significantly more time and annotations. (B) Mean trial duration. The positive instruction framing ($I+$) \emph{increased} duration by $+22$ s, whereas positive valence ($V+$) and provenance labeling ($P+$) \emph{reduced} it by $-33$ s and $-32$ s, respectively. (C) Revision rates (proportion of trials where a participant added a marker, then deleted it). Provenance labeling ($P+$) increased revisions ($+7.0\%$), while positive instruction framing ($I+$) and positive valence ($V+$) decreased them ($-9.6\%$, $-12.4\%$, respectively), indicating that contextual cues modulate annotation confidence and the willingness to revise judgments.}
    \label{fig:engagement}
\end{figure*}

Detection performance was evaluated using window-based metrics derived from the temporal overlap between participant annotations and ground-truth synthetic regions, with a sliding window of $0.2s$ and a boundary tolerance of $\pm200ms$. Each window is assigned a binary ground-truth label and a binary participant decision, giving four outcomes: a window is a true positive (TP) if synthetic and marked, a true negative (TN) if authentic and unmarked, a false positive (FP) if authentic but marked, and a false negative (FN) if synthetic but unmarked. This allows performance to degrade proportionally to the fraction of the audio incorrectly flagged. Accuracy is the proportion of windows for which the participant's decision matched the ground truth. For utterance types containing only one class of windows (e.g., fully authentic or fully synthetic), accuracy reduces to TNR or TPR, respectively, and the complementary metrics are undefined.

Two forms of attention control were implemented to ensure adequate task engagement. First, participants were instructed to listen to each utterance in full before responding, and a compliance criterion of 90\% of all trials was imposed. This threshold was chosen to tolerate occasional accidental clicks or overly rapid responses due to overconfidence, while still discouraging multitasking that could impair performance, given the task’s expected complexity \cite{adler2015effects}. Second, adherence to task instructions was probed with an additional question that appeared on a random one-third of trials, asking participants to indicate which scenario they had been instructed to follow. Participants selected from five options: the two true scenarios ("Monitoring for attacks."/$I-$ and "Evaluating new voice technology."/$I+$), two decoy scenarios ("Creating new synthetic voices." and "Moderating for offensive language."), and a default option ("I did not pay attention"). Because the task scenario remained visible on-screen throughout, submissions were rejected if a participant provided more than one incorrect answer. This defensive design \cite{allahbakhsh2013quality} was used to help ensure that the collected data accurately reflected participants’ actual abilities within a crowdsourcing framework.

Platform interactions were timestamped throughout the task, with an action log recording marker additions, removals, and response selections. Only responses present at the moment of saving were treated as final answers. Participants were debriefed at the end of their session and shown a visualization of their marker placements against ground truth, alongside their evaluation responses.

\subsection{Trust cues}
Trust-influencing manipulations were designed to reflect real-world scenarios where users encounter potentially manipulated media under varying degrees of contextual trust. 

Each participant was randomly assigned to one of two contextual scenarios, presented at the beginning of the session and confirmed through an initialization button ("I understand"). The positive trust scenario framed the task as evaluating a groundbreaking communication technology designed to benefit users. The negative trust scenario warned participants about a nefarious entity deliberately targeting the platform's users. All participants received identical task instructions: they would act as content moderators for a (fictitious) social media platform, and their decisions would affect the platform's users. These instructions remained visible throughout the experiment to maintain the framing context.

For valence priming, each participant was randomly assigned at the start of the session to either a positive or a negative valence condition. 20 images were sampled for each condition (see Section \ref{sec:dataset}), meaning each participant would see a new image each trial, but that all participants assigned to a condition would see the same set of images. Participants viewed images exclusively from their assigned condition, presented in randomized order, before each trial.

Finally, on each individual trial, there was a 50\% probability that the speech excerpt would be accompanied by a provenance label. Unlike standard "AI-generated" disclosures which often function as neutral warnings or negative signals \cite{gamage2025labeling}, our label explicitly affirmed a trusted origin ("This originates from a verified source"). This design choice was motivated by arguments that positive provenance cues may be more effective at building cognitive trust than mere disclosure \cite{harris2024synthetic, fisher2025something}. The label was randomly assigned and had no correlation with whether the audio was genuine or synthetic.

\subsection{Platform}
The experiment was conducted using Prolific\footnote{\url{https://www.prolific.com}}, a web-based platform for recruiting research participants. The decision to

\begin{figure*}[!h]
    \centering
    \includegraphics[width=1\textwidth]{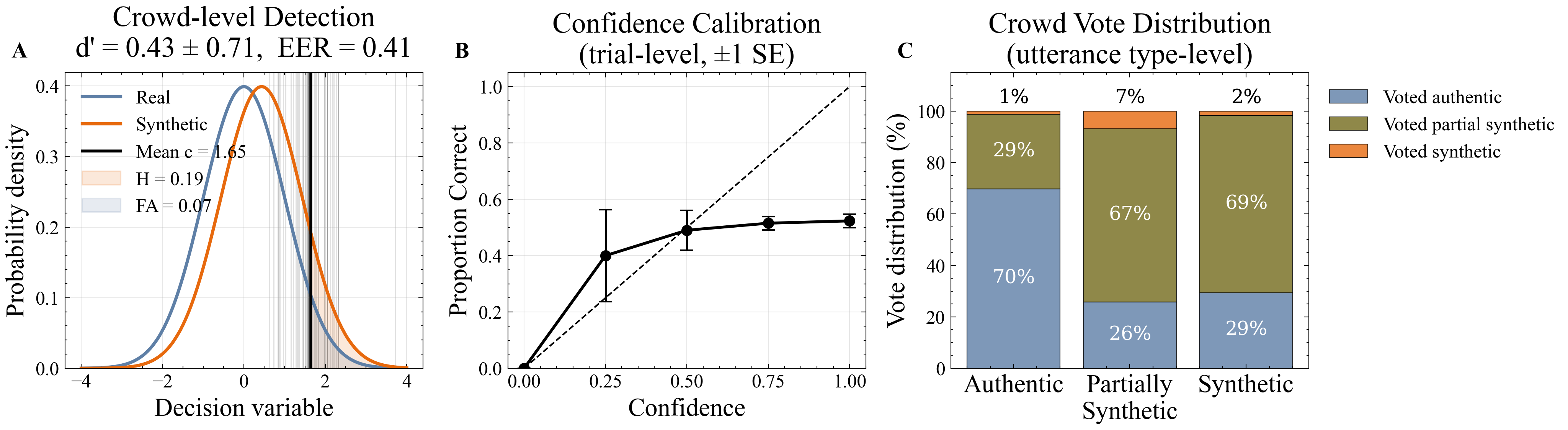}
    \caption{Crowd-level detection performance. (A) Raw discriminability. Distributions show the internal decision variable for real and fake windows, indicating near-chance discrimination. (B) Confidence calibration. Observed accuracy as a function of self-reported confidence indicates that participants were systematically overconfident. (C) Majority-vote accuracy by utterance type. Authentic and partially synthetic clips were correctly identified in 80\% of cases; fully synthetic clips were never correctly classified.}
\label{fig:crowd}
\end{figure*}
\vspace{-0.2cm} 
\noindent use a crowdsourcing platform was driven by the need to recruit participants who shared the same linguistic background as the speech dataset, thereby reducing potential effects from non-native speakers \cite{san2025human}. Submissions were monitored and  approved on a rolling basis until the participant quota was filled. The experimental task was delivered through a custom Python application implemented in Streamlit\footnote{\url{https://streamlit.io}}, a framework for deploying interactive web applications, and was visually designed to evoke the structure of a social media platform (e.g. the priming image presented as the trial thumbnail).

Prior to deployment, an in-house preliminary usability study $(n = 6)$ was conducted to assess interface clarity and ease of use. Participants were familiar with spoofed audio content through prior experience. In response to the feedback, the instructions were set to a retrospective annotation paradigm, allowing participants to listen to the audio as many times as needed. An additional pilot study was conducted via Prolific $(n = 5)$ to validate end-to-end functionality. Data from this pilot were not included in the analysis and instead used to confirm that the task was feasible within the advertised 40-minute average and to verify data integrity in the web deployment format, with the participants being compensated for their time. Pilot completion times averaged 48 minutes, with the longest submission lasting 76 minutes and 16 seconds. Based on this pilot, additional "sanity checks" were implemented to verify participant engagement, namely the attention-to-task instruction check and confirmation that the full utterance was attended to prior to submitting responses, as described in Section \ref{sec:taskdesign}. 

\subsection{Participants}
The study was conducted online in December 2025 and totaled 152 participants who attempted the study. Submissions were manually validated using pre-established sanity checks (see Section \ref{sec:taskdesign}) to ensure adequate engagement with the task. Inclusion criteria included adults aged 18 or older and normal or corrected hearing. To control for linguistic variability, participants were limited to native English speakers from the United States. No specialized technical or linguistic expertise was required, although participants were informed that the study involved listening to utterances and making authenticity judgments. Of the 50 validated submissions, 3 were excluded from analysis due to missing data, yielding a final sample of n=47 (25 male, 21 female; age range $18-55+, \mu=41$). Participants with validated submissions were compensated for their time.

\section{Results}
Each utterance was processed by aligning participant marker placements to ground-truth synthetic regions via sliding window scoring (see Figure \ref{fig:Markers}), producing window-level metrics from which an overall verdict per utterance was derived and inter-participant agreement calculated. The results reveal a critical tension in the socio-technical trust framework: while the environmental context was hypothesized to modulate detection sensitivity, utterance type overwhelmingly dictated outcomes. 

\vspace{-0.35cm}
\begin{table}[h!]
\centering
\resizebox{\columnwidth}{!}{%
\begin{tabular}{lcccccc}
\textbf{\begin{tabular}[c]{@{}l@{}}Utterance Type/ \\ Window Type\end{tabular}} &
  \textbf{Accuracy} &
  \textbf{\begin{tabular}[c]{@{}c@{}}Majority \\ Vote\end{tabular}} &
  \textbf{TPR} &
  \textbf{FPR} &
  \textbf{TNR} &
  \textbf{FNR} \\ \hline
Authentic               & 96.4\% & 4/5  & n/a  & 3.6\%  & 96.4\% & n/a      \\
Synthetic               & 8.3\%  & 0/5  & 8.3\%  & n/a  & n/a   & 91.7\% \\
Partially Synth. - Fake & 29.3\% & n/a   & 29.3\% & n/a      & n/a& 70.7\% \\
Partially Synth. - Real & 88.6\% & n/a    & n/a   & 11.4\% & 88.6\% & n/a      \\
Partially Synthetic     & 56.1\% & 8/10 & 29.3\% & 11.4\% & 88.6\% & 70.7\% \\ \hline
\textbf{Overall} &
  \textbf{55.8\%} &
  \textbf{12/20} &
  \textbf{19.4\%} &
  \textbf{7.1\%} &
  \textbf{92.9\%} &
  \textbf{80.6\%}
\end{tabular}%
}
\caption{Window- and trial-level detection performance by utterance type. Fully synthetic speech was never correctly identified at the trial level (0/5).}
\label{tab:results}
\end{table}
\vspace{-0.6cm}

\begin{figure}[H]
    \centering
    \includegraphics[width=1\linewidth]{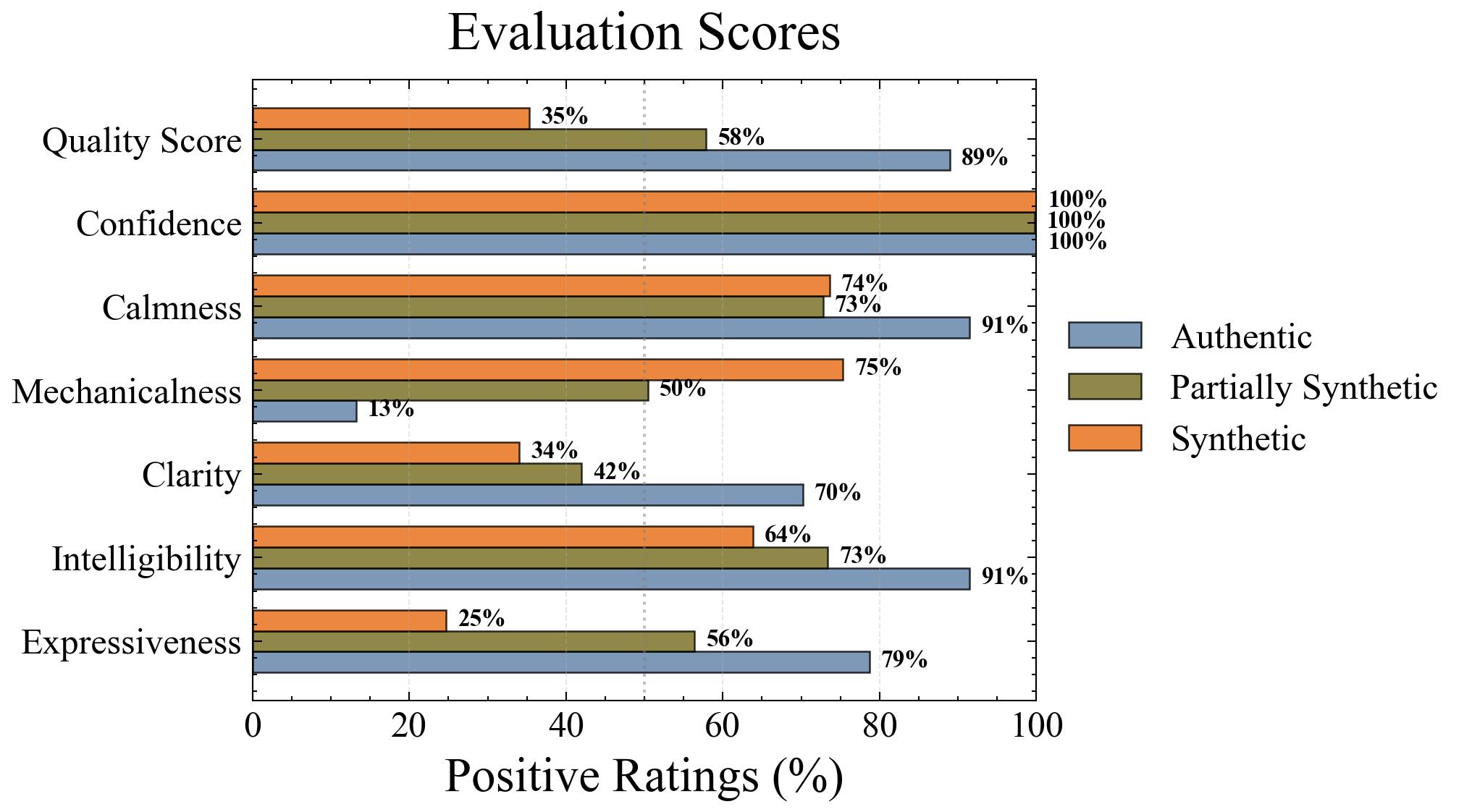}
    \caption{Percentage of positive ratings (\textit{"Agree", "Strongly Agree"}). Authentic speech consistently received the highest subjective ratings across all dimensions, while fully synthetic speech received the lowest. Note that mechanicalness is presented as rated (higher = more mechanical), such that lower scores indicate higher perceived quality for this dimension.}
\label{fig:perceptual_eval}
\end{figure}

\noindent \textbf{Detection performance}: Linear mixed-effects models, estimated via maximum likelihood using the Python \texttt{statsmodels} package \cite{seabold2010statsmodels},  revealed a strong ordinal effect of utterance type on detection accuracy (\textbf{H1} supported), with authentic speech yielding highest accuracy and synthetic lowest. The same ordinal pattern held for perceptual quality ratings (\textbf{H2} supported): authentic utterances received the highest quality scores, followed by partially synthetic, with fully synthetic rated lowest (see Figure \ref{fig:perceptual_eval}). 

\noindent \textbf{Trust cue effects}: Trust cues exerted no reliable main effects on either detection accuracy or quality evaluation (\textbf{H1a} and \textbf{H2a} not supported). Influence was confined to interaction terms, suggesting that contextual framing modulates responses only in combination with specific audio or participant characteristics rather than shifting detection performance globally.
\noindent \textbf{Collective decision-making}: aggregating window-level markings  revealed low crowd discriminability ($d' = 0.43 \pm 0.71$, EER $= 41\%$,  see Figure \ref{fig:crowd}A), with substantial between-person variability. At the trial level, majority voting classified trials as follows:
\vspace{-0.3cm}
\begin{equation}
\text{MajorityVote}(\text{trial}) = 
\begin{cases} 
\text{authentic}           & \text{if } c = 0 \\
\text{partially synthetic} & \text{if } 0 < c \leq 70\% \\
\text{fully synthetic}     & \text{if } c > 70\%
\end{cases}
\end{equation}
where $c$ denotes coverage (proportion of utterance duration covered by participant-added markers). Trial-level accuracy was 60\% (12/20 correct), with average reported confidence of 71.5\%. Performance varied substantially by ground-truth utterance type (see Table \ref{tab:results}, \textit{Majority Vote}): authentic utterances achieved 80\% accuracy (4/5 correct) and partially synthetic 80\% (8/10 correct), but fully synthetic speech was \emph{never} correctly classified (0/5 correct). Instead, it was systematically misclassified as partially synthetic in all cases, indicating that participants detected manipulation but consistently underestimated its extent.

\vspace{-0.2cm}
\begin{table}[h!]
\centering
\small
\resizebox{\columnwidth}{!}{
\begin{tabular}{llrrrl}
\hline
\textbf{Hypothesis} & \textbf{Predictor} & \textbf{Coef} & \textbf{SE} & \textbf{p (adj)} & \textbf{Effect size} \\ \hline
\multirow{2}{*}{H1: Accuracy} & \begin{tabular}[c]{@{}l@{}}Synth. vs \\ Auth.\end{tabular} & -0.871 & 0.020 & $<$.001*** & \begin{tabular}[c]{@{}l@{}}Cohen's $d$=-4.04, \\ $R^2$=.67\end{tabular} \\
 & \begin{tabular}[c]{@{}l@{}}Part. vs \\ Auth.\end{tabular} & -0.402 & 0.017 & $<$.001*** & \begin{tabular}[c]{@{}l@{}}Cohen's $d$=-1.87, \\ $R^2$=.37\end{tabular} \\ \hline
\multirow{4}{*}{H1a: TPR} & \begin{tabular}[c]{@{}l@{}}Synth. vs \\ Part.\end{tabular} & -0.191 & 0.023 & $<$.001*** & \begin{tabular}[c]{@{}l@{}}Cohen's $d$=-0.67, \\ $R^2$=.09\end{tabular} \\
 & $I-$ & -0.019 & 0.048 & .750 & Cohen's $d$=-0.07 \\
 & $V-$ & 0.040 & 0.046 & .495 & Cohen's $d$=0.14 \\
 & $P-$ & 0.031 & 0.022 & .258 & Cohen's $d$=0.11 \\ \hline
\multirow{2}{*}{\begin{tabular}[c]{@{}l@{}}H2: Quality\\ Score\end{tabular}} & \begin{tabular}[c]{@{}l@{}}Synth. vs \\ Auth.\end{tabular} & -1.154 & 0.070 & $<$.001*** & \begin{tabular}[c]{@{}l@{}}Cohen's $d$=-1.53, \\ $R^2$=.23\end{tabular} \\
 & \begin{tabular}[c]{@{}l@{}}Part. vs \\ Auth.\end{tabular} & -0.716 & 0.060 & $<$.001*** & \begin{tabular}[c]{@{}l@{}}Cohen's $d$=-0.95, \\ $R^2$=.13\end{tabular} \\ \hline
\multirow{5}{*}{\begin{tabular}[c]{@{}l@{}}H2a: Quality\\ Score\end{tabular}} & \begin{tabular}[c]{@{}l@{}}Synth. vs \\ Auth.\end{tabular} & -1.154 & 0.070 & $<$.001*** & \begin{tabular}[c]{@{}l@{}}Cohen's $d$=-1.53, \\ $R^2$=.23\end{tabular} \\
 & \begin{tabular}[c]{@{}l@{}}Part. vs \\ Auth.\end{tabular} & -0.717 & 0.060 & $<$.001*** & \begin{tabular}[c]{@{}l@{}}Cohen's $d$=-0.95, \\ $R^2$=.13\end{tabular} \\
 & $I-$ & -0.137 & 0.117 & .350 & Cohen's $d$=-0.18 \\
 & $V-$ & 0.025 & 0.112 & .824 & Cohen's $d$=0.03 \\
 & $P-$ & -0.038 & 0.050 & .534 & Cohen's $d$=-0.05 \\ \hline
\end{tabular}}
\caption{Linear mixed model results, $N=940$ trials, 47 participants, 20 trials each. All models include a random intercept per participant estimated via maximum likelihood. H1 and H2 show strong effects of utterance type on both detection accuracy and perceived quality. H1a and H2a show no significant effect.}
\label{tab:lmm-results}
\end{table}
\vspace{-0.5cm}

\section{Discussion}
Utterance type emerged as the dominant determinant of detection, suggesting that the environmental dimension of trust may fall to a secondary role in real-time detection tasks. This pattern may reflect a mismatch in listener expectations: participants entered the task expecting robotic or background artifacts \cite{warren2024better,amirkhani2025detecting}, yet encountered realistic soundscapes and utterances with only subtle word-level manipulations. Such expectation violations could have pulled attention away from contextual cues toward acoustic scrutiny. Evaluative judgments closely tracked audio realism, suggesting that listeners build an internal representational ranking for quality (authentic $>$ partially synthetic $>$ synthetic) that is accessible to conscious evaluation but does not transfer to overt detection decisions. This implicit-explicit dissociation has direct repercussions: attackers may successfully sway into compliance listeners who nonetheless feel something is "off", without that unease surfacing as an actionable response. Participants' engagement metrics offer additional insight into their decision-making patterns (see Figure \ref{fig:engagement}). Participants spending the least time on authentic trials suggests that they can readily recognize real content but are poor at detecting synthetic material. When uncertain, they draw on other information sources (e.g., the trust-related changes in trial duration and revision rates), yet they still perform poorly overall on the task. The combination of below-chance accuracy, overconfidence, and negligible contextual benefits indicates that listeners cannot reliably serve as sole gatekeepers against synthetic speech. 

\section{Limitations and future directions}
Exploring context-driven human decision-making factors is a multidimensional problem. If automated detection studies test the strength of countermeasures \cite{ alali2025partial, liu2023asvspoof, wang2024asvspoof, li2025we,  khan2023battling}, with human listener studies exploring themes of detection \cite{warren2024better,  alali2025partial, amirkhani2025detecting,  san2025human}, our study tackled the environment dimension by integrating variables that would otherwise be discarded in protocol development. Containing these factors within a practical platform constrained ecological validity, especially as the explicit task framing primed participants for suspicion and thus likely attenuating the trust cue effects. In naturalistic settings, where synthetic speech is encountered without warning, people’s behavior and building of trust would be additionally influenced by their surroundings, disposition, or prior experience with the platform \cite{scott2020human, inproceedingstrust}. Research on synthetic speech detection would benefit from in-the-wild designs that study participants who are unaware of its potential presence. More broadly, as acoustic quality continues to improve, research attention should shift from detection performance toward identifying specific risk scenarios that can inform regulatory deterrents \cite{bhatti2026can}. Finally, the stimuli set of 20 utterances limits the generalizability of these findings across the broader landscape of synthesis systems and speaker identities. A beneficial extension of this paradigm would be connecting to ongoing research into speaker trustworthiness and how that is embedded into synthetic speech \cite{torre2018trust, maltezou2025voice}.

\vspace{-0.2cm}
\vspace{-0.3cm}

\section{Conclusion}
This study challenges the "one solution fits all" approach to synthetic speech detection by investigating (partial) voice deepfake detection as a \emph{socio-technical process shaped by trust}. Although participants expressed confidence in their judgments, they were generally unable to detect synthetic speech, with detection performance improving only when authentic speech served as an acoustic anchor for suspicion as in the case of partial synthetic utterances. Contextual trust cues, intended to modulate suspicion, did not significantly alter detection outcomes. As such, it seems that the unreliable, naive listener, instead of being supported by additional information, may experience cognitive overload rather than improved accuracy. More broadly, if moving towards ecological validity itself drives complexity beyond what naive listeners can manage, the expectation of routine human detection in an increasingly synthetic media landscape warrants serious reconsideration.

If these results generalize, the amount of synthetic speech that passes the perceptual threshold is likely far greater than reported, and more so as people engage with it as they would with authentic speech. If this entanglement of real and synthetic continues, the assignment of truth will become increasingly fuzzy. Using automated decision systems as support will then have to be tailored to specific user needs and affordances.

The path forward seems to point toward cooperative architectures that distribute detection across both human judgment and technical countermeasures. Such joint systems should handle uncertainty rather than demand binary verdicts, and account for the overconfidence and response biases, some of which are documented here. Building such systems will require a clearer understanding of the conditions under which people reason, doubt, and decide, as well as how the platforms they frequent can be designed to support, rather than undermine, that process.

\section{Acknowledgments}
This work was carried out as part of the VoCS (Voice in Communication Sciences) doctoral network, funded by the European Union's Horizon Europe Framework programme under Grant Agreement No 101168998 and partially supported by the Academy of Finland (Decision No. 349605, project ”SPEECHFAKES”). This study was submitted to and received no objection from the ethics committee of the University of Eastern Finland. We thank the reviewers for their constructive feedback, and the participants and volunteers whose involvement made this work possible.

\bibliographystyle{IEEEtran}
\bibliography{references}

\end{document}